

\input harvmac
\noblackbox
%
%

%
%

\def\Title#1#2{\ifx\answ\bigans \nopagenumbers
\abstractfont\hsize=\hstitle\rightline{#1}%
\vskip .5in\centerline{\titlefont #2}\abstractfont\vskip .5in\pageno=0
\else \rightline{#1}
\vskip .8in\centerline{\titlefont #2}
\vskip .5in\pageno=1\fi}
\ifx\answ\bigans

\else

 \font\absi=cmmi10 scaled\magstep1
\font\absis=cmmi7 scaled\magstep1 \font\absiss=cmmi5 scaled\magstep1
\font\abssy=cmsy10 scaled\magstep1 \font\abssys=cmsy7 scaled\magstep1
\font\abssyss=cmsy5 scaled\magstep1 
\skewchar\absi='177 \skewchar\absis='177 \skewchar\absiss='177
\skewchar\abssy='60 \skewchar\abssys='60 \skewchar\abssyss='60
\fi
%
%

\def\hf{{1\over2}}
\def\sq{{\vbox {\hrule height 0.6pt\hbox{\vrule width 0.6pt\hskip 3pt
   \vbox{\vskip 6pt}\hskip 3pt \vrule width 0.6pt}\hrule height 0.6pt}}}
\def\ajou#1&#2(#3){\ \sl#1\bf#2\rm(19#3)}
\def\apm{{\alpha^\prime}}

\def\frac#1#2{{#1 \over #2}}

\def\etp{e^{2\phi}}
\def\eps{\epsilon}
\def\mn{{\mu\nu}}
\def\ls{{\lambda\sigma}}
\def\tr{{\rm tr}}
\def\Tr{{\rm Tr}}
\def\vx{{\vec x}}
\def\rs{{\overline{r}}}
%
%
\lref\sennew{A. Sen, {\sl Quantization of Dyon Charge and Electric-Magnetic
Duality in String Theory}, Tata Institute preprint TIFR-TH-92-46
(hep-th/9209016) (September 1992);
{\sl Electric Magnetic Duality in String Theory}, Tata Institute preprint
TIFR-TH-92-41 (hep-th/9207053) (July 1992).}
%
\lref\HL{J. A. Harvey and J. Liu, Phys. Lett. {\bf B268} (1991) 40.}
\lref\BaVi{M. Barriola and A. Vilenkin, Phys. Rev. Lett. {\bf 63} (1989) 341.}
\lref\harr{B. Harrington and H. Shepard, Phys. Rev. {\bf D17} (1978) 2122.}
\lref\rossi{P. Rossi, Nucl. Phys. {\bf B149} (1979) 170.}
\lref\wenwit{X. G. Wen and E. Witten, Nucl. Phys. {\bf B261} (1985) 651.}
\lref\sorkin{R. D. Sorkin, Phys. Rev. Lett. {\bf 51} (1983) 87.}
\lref\gperry{D. J. Gross and M. J. Perry, Nucl. Phys. {\bf B226} (1983) 29.}
\lref\banksetal{T. Banks, M. Dine, H. Dijkstra and W. Fischler,
Phys. Lett. {\bf B212} (1988) 45.}
\lref\rohmwitt{R. Rohm and E. Witten, Ann. Phys. (NY) {\bf 170} (1986) 454.}
\lref\sss{C. G. Callan, J. A. Harvey and A. Strominger, {\sl Supersymmetric
String Solitons}, in {\it String Theory and Quantum Gravity '91: Proceedings
of the Trieste Spring School \& Workshop}, pp. 208--244,
World Scientific, Singapore (1991).}
\lref\fivenon{C. G. Callan, J. A. Harvey and A. Strominger, Nucl. Phys.
{\bf B367} (1991) 60.}
\lref\josh{J. A. Frieman and C. T. Hill, {\sl Imploding Monopoles},
SLAC preprint SLAC-PUB-4283 (1987).}
\lref\hetsol{A. Strominger, Nucl. Phys. {\bf B343} (1990) 167;
E: Nucl. Phys. {\bf B353} (1991) 565.}
\lref\MOOL{C. Montonen and D. Olive, Phys. Lett. {\bf 72B} (1977) 117. }
\lref\osborn{H. Osborn, Phys. Lett. {\bf 83B} (1979) 321.}
\lref\khurione{R. R. Khuri, {\sl A Multimonopole Solution in String Theory},
Texas A\&M preprint CTP/TAMU-33/92 (hep-th/9205051) (April 1992);
{\sl A Heterotic Multimonopole Solution}, Texas A\&M preprint
CTP/TAMU-35/92 (hep-th/9205081) (April 1992).}
\lref\BEDR{E. A. Bergshoeff and M. de Roo, Nucl. Phys. {\bf B328} (1989) 439.}
\lref\world{C. G. Callan, J. A. Harvey and A. Strominger, Nucl. Phys.
{\bf B359} (1991) 611.}
\lref\raj{R. Rajaraman, {\it Solitons and Instantons}, North-Holland,
Amsterdam (1982).}
\lref\gpy{D. J. Gross, R. D. Pisarski and L. G. Yaffe, Rev. Mod. Phys.
{\bf 53} (1981) 43.}

%
%
\Title{\vbox{\baselineskip12pt
\hbox{EFI-92-67}
\hbox{IFP-434-UNC}
\hbox{hep-th/9211056}}}
{Magnetic Monopoles in String Theory}
{
\baselineskip=12pt
\bigskip
\centerline{Jerome P. Gauntlett, Jeffrey A. Harvey}
\bigskip
\centerline{\sl Enrico Fermi Institute, University of Chicago}
\centerline{\sl 5640 Ellis Avenue, Chicago, IL 60637 }
\centerline{\it Internet: jerome@yukawa.uchicago.edu}
\centerline{\it Internet: harvey@poincare.uchicago.edu}
\bigskip
\centerline{James T. Liu}
\bigskip
\centerline{\sl Institute of Field Physics}
\centerline{\sl Department of Physics and Astronomy}
\centerline{\sl University of North Carolina}
\centerline{\sl Chapel Hill, NC 27599-3255}
\centerline{\it Internet: jtliu@physics.unc.edu}
\medskip

\bigskip
\centerline{\bf Abstract}
Magnetic monopole solutions to heterotic string theory are discussed in
toroidal compactifications to four spacetime dimensions.  Particular
emphasis is placed on the relation to previously studied fivebrane
solutions in ten dimensions and on the possibility of constructing
exact monopole solutions related to symmetric fivebranes.

}


\Date{11/92}

%
%

\newsec{Introduction}
Although a wide variety of approximate and exact soliton solutions to
string theory are now known, many of the most important questions
involving solitons in string theory are still open. These include the
proper treatment of collective coordinates, and the possibility of a
strong-weak coupling duality in string theory \hetsol\ modeled after
the conjectured electric-magnetic duality in $N=4$ gauge theory
\refs{\MOOL,\osborn}.  In this regard magnetic monopoles provide a
particularly important subset of possible soliton solutions to string
theory. A number of approximate monopole solutions have already been
studied. Recently it has been claimed that there are also monopole
solutions which provide exact solutions of string theory \khurione.
This work was motivated by the desire to better understand these
solutions.  We will begin in section 2 with a quick summary of previous
work on monopole solutions to string theory. In section 3 we will
discuss the relation between fivebrane solutions and magnetic
monopoles. We show that two previously known monopole solutions can be
constructed from a periodic array of ``gauge'' and ``neutral"
fivebranes, respectively.  We then to turn to the construction of
solutions corresponding to a periodic array of ``symmetric''
fivebranes. In section 4 we develop the properties of these solutions
and compare our results with previous work. We end with our conclusions
in section 5.

\newsec{Summary of Previous Work}

There are a number of possible magnetic monopole solutions to string
theory compactified from ten to four dimensions. First consider
standard compactifications of heterotic string theory on a Calabi-Yau
space $K$ with $H^1(K)=0$ and gauge symmetry breaking by Wilson lines.
If the unbroken gauge group has a $U(1)$ factor one might expect
magnetic monopole solutions to exist. Demanding that the asymptotic
form of the $U(1)$ gauge field is that of a magnetic monopole, one asks
if the configuration can be extended smoothly over the whole of space.
It turns out that this condition places  some rather subtle topological
restrictions on the allowed magnetic charges \wenwit.  Within each
topological class satisfying these restrictions we would expect that
there is a magnetic monopole solution to string theory.  However, not
much is known about the detailed form of the solution since it involves
massive Kaluza-Klein states in a non-trivial way.

If we consider less realistic compactifications there are several new
possibilities for monopole solutions. In particular suppose that the
compact space has the form $S^1 \times K'$ where $K'$ is arbitrary as
long as it provides a solution to the string equations of motion. Then
the low-energy four-dimensional theory will have (at least) two $U(1)$
gauge fields coming from the components $g_{\mu 4}$ of the metric and
$B_{\mu 4}$ of the antisymmetric tensor field. Here $\mu$ is a
spacetime index and $x^4$ is the coordinate along the $S^1$. At the
level of low-energy field theory there are magnetic monopole solutions
carrying magnetic charge under both of these $U(1)$ factors. At lowest
order the ``metric'' magnetic monopole is just that constructed by
Sorkin \sorkin\ and by Gross and Perry \gperry. In \banksetal\ it was
argued that starting from this solution one can construct a solution to
string theory to all orders in perturbation theory in the parameter
$\alpha' /R^2$ with $R$ the radius of the $S^1$.  By utilizing the
string duality $R \rightarrow \alpha' /R$ a solution involving
$B_{\mu 4}$ was also constructed in \banksetal.  Monopole solutions of
heterotic string theory involving $B_{\mu 4}$ were discussed in a
general context in \rohmwitt\ where it was emphasized that the gauge
invariant field strength is given by the antisymmetric three-form
$H_{(4)\mn}\equiv H_{4\mn}$ and that the Yang-Mills Chern-Simons
contributions can play
an important role. We will elaborate on this point later.

Finally there exist a class of magnetic monopole solutions which are
perhaps most closely related to the usual 't~Hooft-Polyakov monopole.
These occur when a non-abelian gauge symmetry of the four-dimensional
field theory is spontaneously broken by a light Higgs field. Two
closely related examples have been studied.  In the first example we
consider the theory close to the self-dual radius of the $S^1$.  At the
self-dual point there is a well-known $SU(2)$ gauge symmetry and a set
of massless scalars in the triplet representation of $SU(2)$.  As one
moves away from this point the $SU(2)$ is broken to $U(1)$ and we
expect monopole solutions. This solution was constructed in
\banksetal\ in a perturbative expansion away from the self-dual point.
A very similar situation occurs in a toroidal compactification of
arbitrary radius down to four dimensions ($K'=T^5$) in which case one
has $N=4$ supersymmetry and the gauge supermultiplets contain scalars
in the adjoint representation of the gauge group. Since a potential for
these scalars is forbidden by supersymmetry, one can assign arbitrary
vacuum expectation values for these fields again leading to symmetry
breaking with $U(1)$ factors in the unbroken gauge group. The
corresponding monopoles were constructed in \HL\ where it was shown
that the monopoles preserve half of the supersymmetries and saturate a
Bogomol'nyi bound.

Although all of these monopoles can presumably be extended to solutions
to string theory, they all receive corrections to higher orders in
$\alpha'$. This makes their description in terms of conformal field
theory problematic. It would be very nice to have solutions which are
exact without any higher order corrections.  Such solutions have been
proposed in \khurione\ based on ``symmetric'' fivebrane solutions
\world.  We will argue that the construction of \khurione\ does in
fact lead to exact monopole solutions, although there are some crucial
differences in interpretation between this work and \khurione. In
addition, similar constructions allow us to relate aspects of the
solutions of \banksetal\ and \HL\ to the ``neutral'' and ``gauge''
fivebranes.

\newsec{Fivebranes and Monopoles}

A fivebrane is an extended soliton solution to 10 dimensional string
theory with 5+1 dimensional translational symmetry. Explicit fivebrane
solutions have been constructed from a generalization of Yang-Mills
instantons in which the four-dimensional instanton sits in the directions
transverse to the fivebrane.  When such objects are compactified to
four dimensions, they can be classified by the embedding of the core
instanton in space-time and internal space \hetsol.  In particular, if
the instanton lies in 3 space directions and 1 internal direction, it
appears as a particle from the four dimensional point of view. In the
following, we show that these pointlike solitons can be further
identified as magnetic monopoles.

\subsec{Gauge, neutral and symmetric fivebranes}

The fivebrane solutions are constructed from the low-energy effective
action for the massless fields of the heterotic string.  At lowest
order in $\apm$, the effective action is given by  $N=1$ super
Yang-Mills coupled to supergravity theory.  In ``sigma model''
variables, the bosonic part is
\eqn\sYM{S={1\over 2\kappa^2}\int d^{10}x\sqrt{g}e^{-2\phi}\left[
R+4(\partial\phi)^2-{1\over3}H^2-{\apm\over30}\Tr F^2\right]}
where the Yang-Mills gauge fields are in the adjoint representation of
$E_8\times E_8$ or $SO(32)$ with the trace conventionally normalized so
that $\tr(t^at^b)=\delta^{ab}$ in the fundamental representation.

The $H$ field is given at lowest order by
$H=dB+\apm(\omega_3^L-{1\over30}\omega_3^{YM})$ with $\omega_3^L$ and
$\omega_3^{YM}$ being the Lorentz and Yang-Mills Chern-Simons three-forms
respectively.  Without the Lorentz Chern-Simons form this action
has a well known supersymmetric completion. When this term is included
one must keep including additional terms in a power series in $\apm$ in
order to maintain supersymmetry.  In carrying out this procedure it is
found that the generalized spin connections
$\Omega_{\pm M}^{AB}\equiv\omega_M{}^{AB} \pm H_M{}^{AB}$ play a
central role \BEDR.  In particular, one finds that the three-form $H$
is recursively defined by evaluating the Lorentz Chern-Simons term in
the definition of $H$ above using the generalized connection
$\Omega_+$. Thus the Bianchi identity that supplements \sYM\ is most
informatively written as
\eqn\bianc{dH=\apm ( \tr R(\Omega_+)\wedge R(\Omega_+)
- {1\over30} \Tr F\wedge F)+O(\apm^2).}
The role of the generalized connections can also be understood from the
sigma-model point of view. For a discussion of this in the context of
fivebrane solutions see \sss.

The supersymmetry transformations for the fermionic fields are, to
lowest order, given by
\eqn\fbsusy{\eqalign{
\delta\chi&=F_{MN}\gamma^{MN}\epsilon\cr
\delta\lambda&=\left(\gamma^M\partial_M\phi-{1\over 6}H_{MNP}\gamma^{MNP}
\right)\epsilon\cr
\delta\psi_M&=\left(\partial_M+{1\over 4}{\Omega^{AB}_{-M}}\gamma_{AB}
\right)\epsilon.\cr
}}

The fivebrane ansatz preserves a chiral half of the supersymmetries and
is given by \refs{\hetsol,\world}
\eqn\fivebr{\eqalign{
F_\mn &=  \pm\hf\eps_\mn{}^\ls F_{\ls}\cr
H_{\mn\lambda} &= \mp\eps_{\mn\lambda}{}^\sigma\partial_\sigma\phi\cr
g_\mn& =\etp\delta_\mn, \qquad g_{ab}=\eta_{ab}\cr}}
where $\mu, \nu, \ldots$ denote transverse space, and $a, b, \ldots$
orthogonal space indices.  For the gauge part of the solution, the
first line indicates that $F_\mn$ is an
{\hbox{(anti-)}\penalty\exhyphenpenalty}self-dual field
strength, and in particular can be solved by an instanton configuration
in an $SU(2)$ subgroup of the gauge group.

{}From this starting point, there are two approaches to constructing
the rest of the fivebrane solution.  The first is to solve the Bianchi
identity \bianc\
%
%
perturbatively in $\apm$.  Starting with a $F_\mn=O(1)$ instanton
solution, the Bianchi identity tells us that the dilaton and hence the
curvature is $O(\apm)$, so to lowest order, we can drop the $R\wedge R$
term to give
\eqn\gsol{
\nabla_\rho\nabla^\rho \phi = \mp {\apm\over 120} \eps^{\mn\ls}
\Tr F_\mn F_\ls}
which can be solved for a given multi-instanton configuration.  For the
charge one self-dual instanton of scale size $\rho$ one obtains
\eqn\gexp{e^{2 \phi} = e^{2 \phi_0} + 8 \apm {(x^2 + 2 \rho^2) \over
                        (x^2 + \rho^2)^2}.}
These fivebrane solutions are referred to as gauge fivebranes
\refs{\hetsol,\world} and receive higher order corrections in $\apm$.
Nevertheless, it is possible to maintain supersymmetry  and construct a
solution order by order in $\apm$ using non-renormalization arguments
based on the six-dimensional symmetry of the low-energy fivebrane
effective action \fivenon.

The neutral fivebrane solution is obtained from the gauge fivebrane
\gexp\ by taking the limit $\rho\to 0$ to get
\eqn\nexp{e^{2 \phi} = e^{2 \phi_0} +  {n\apm \over x^2}.}
Although only the solution with $n=8$ is obtained in this limit the
solution exists for all positive integers $n$\foot{If $n$ is negative
the dilaton becomes imaginary when $x^2<-n\apm e^{-2 \phi_0}$ and
the physical interpretation of the solution is unclear.}.
The most general neutral multi-fivebrane
configuration is obtained by solving $\sq\,\etp=0$ assuming $S^3$
symmetry, giving
\eqn\mnexp{e^{2 \phi} = e^{2 \phi_0} +  \sum_I{n_I\apm  \over
                        (x-x_I)^2}}
for some positive
integers $n_I$ and where $x_I$ denote the locations of the
fivebranes.  Like the gauge fivebranes, the neutral solutions will also
receive higher order corrections in $\apm$.

The second approach to completing the fivebrane solution is, in analogy
with Calabi-Yau compactifications, to embed the generalized spin
connection in the gauge group, $\Omega_+=A$, so that $dH$ vanishes to
all orders in $\apm$.  The condition for this to hold is simply
$\sq\,\etp=0$, just as in the neutral solution \mnexp. In this case
$\Omega_+$, using the above metric ansatz, is given by
\eqn\omegax{\Omega^{mn}_{+\mu}={\sigma_{\mu\nu}}^{mn}\del^\nu2\phi}
with
\eqn\sig{{\sigma_{\mu\nu}}^{mn}={\delta_{\mu\nu}}^{mn}\mp{1\over 2}
{\epsilon_{\mu\nu}}^{mn}}
being anti-self-dual (self-dual) in both pairs of indices. Using the
condition $\sq\,\etp=0$, this implies that $\Omega_+$ is an
(anti-)self-dual $SU(2)$ connection (embedded in $SO(4)$), ensuring the
consistency of equating it with an (anti-)self-dual Yang-Mills
connection (restricted to be in the form of the 't~Hooft ansatz,
discussed below).  The solution with the dilaton given by \nexp\ and
the instanton size given by $\rho=e^{- \phi_0}\sqrt{n\apm}$ is known
as the symmetric fivebrane \world. It is an exact solution of string
theory without higher order corrections as can be seen from various
points of view, including construction of the explicit underlying
superconformal field theory \sss. When the dilaton is given
by \mnexp\ and the instanton sizes are given by
$\rho_I=e^{- \phi_0}\sqrt{n_I\apm}$
we have a multi symmetric fivebrane configuration.

To conclude this review of the fivebrane solutions we note that we have
taken into account the quantization condition on the three-form $H$
that is required for the consistent propagation of strings in this
background \rohmwitt.  Specifically, $H$ must satisfy
\eqn\hq{Q=-{1\over 2\pi^2\apm}\int_{M}H,\qquad Q\in Z}
where the integral is over an arbitrary closed three manifold $M$.  All
of the fivebrane solutions satisfy this condition with $Q=8$ for the
gauge fivebrane and $Q=\sum_I{n_I}$ for the neutral and symmetric
multi-fivebrane solutions. The ``anti-fivebrane" solutions are obtained
using the lower sign in \fivebr\ and although the form of the dilaton
is the same as for the fivebranes (upper sign),
they have opposite $H$-charge $Q$.

\subsec{Periodic instantons and monopoles}

In constructing magnetic monopole solutions from fivebranes we will
exploit a small generalization of the relation between Yang-Mills
instantons and monopoles which we will now review.  For an $SU(2)$
connection, a general $N$ instanton configuration is described by
$8N-3$ parameters: the positions, sizes and relative $SU(2)$ angles of
the $N$ instantons.  The 't~Hooft ansatz gives an explicit $5N$
parameter multi-instanton solution in which all instantons have
identical gauge orientations (see, for example, \raj).
The ansatz involves writing the $SU(2)$ gauge field as
\eqn\thooft{A_\mu(x)=\overline{\Sigma}_\mn\nabla^\nu\ln f(x)}
where the matrix valued 't~Hooft tensor, $\overline{\Sigma}_\mn$, is
antisymmetric and anti-self-dual%
\foot{Note that $\sigma_{\mu\nu}$ defined in \sig\ is a possible choice
for $\overline{\Sigma}_\mn$ if one interprets it as the doublet
representation of $SU(2)$ having been embedded in the fundamental
representation of $SO(4)$.}.
%
%

The self-duality condition then becomes $f^{-1}\sq f=0$, which can be
solved (assuming $S^3$ symmetry) to give
\eqn\multii{f(x)=\sum_{I=1}^{N+1}{\rho_I^2\over(x-x_I)^2}.}
This form of the solution most directly exhibits the conformal symmetry
of the solution. If we take the limit $\rho_{N+1} \rightarrow \infty$
and $x_{N+1} \rightarrow \infty$ with $\rho_{N+1}/x_{N+1}=1$ we obtain
the perhaps more familiar form of the solution
\eqn\multip{f(x)=1+\sum_{I=1}^N{\rho_I^2\over(x-x_I)^2}}
where $x_I$ and $\rho_I$ can now be interpreted as the position and
size of the $I^{\rm th}$ instanton.

To see how monopoles arise from periodic instantons, we consider making
one of the four transverse coordinates ({\it e.g.}~$x^4$) periodic with
period $2\pi R$ and look for solutions to the self-dual equation on the
space $R^3 \times S^1$.

We are thus interested in instanton solutions which are periodic in
$x^4$.  In general, a single periodic instanton can be constructed by
taking an infinite string of identical (up to a gauge transformation)
instantons lined up in the compact direction with spacing $2\pi R$.
Starting with the  't Hooft ansatz \multip\ for instantons with
identical gauge orientation and performing the sum to enforce the
periodicity $x^4\equiv x^4+2\pi R$ gives a single
periodic instanton \harr:
\eqn\singlem{\eqalign{f^{(1)}(\vx, x^4) &= 1 + \sum_{k=-\infty}^\infty
{\rho^2 \over r^2 + (x^4-x^4_0 + 2 \pi k R)^2 }\cr
&= 1 +{\rho^2\over2Rr}\sinh{r\over R}
\bigg/\left(\cosh{r\over R}-\cos{x^4-x^4_0\over R}\right)}}
with $r=| \vec x - \vec x_0|$ and $(\vec x_0,x^4_0)$ the
location of the instanton.

A periodic multi-instanton can be constructed in the same way by
starting with $n$ such strings. This gives
\eqn\multim{f^{(n)}(\vx,
x^4)=1+\sum_{I=1}^n{\rho_I^2\over2Rr_I}\sinh{r_I\over R}
\bigg/\left(\cosh{r_I\over R}-\cos{x^4-x^4_I\over R}\right)}
where $r_I=|\vx-\vx_I|$ is the 3-radius.
The $n$ periodic instantons are located at $(\vx_I, x^4_I)$.

While an uncompactified instanton has a single scale, $\rho$, a single
periodic instanton has two relevant scale parameters, the original
instanton size and the compactification radius of the $S^1$.  The
behavior of the non-abelian field strength of the periodic instanton
depends on the ratio of these scales.  For a single periodic instanton
(centered at the origin) in the asymptotic limit $r\gg R$,
\multim\ reduces to
\eqn\multlim{f=1+{\rho^2\over2Rr}+O(e^{-r/R}).}
{}From \thooft, we see that $A_\mu\sim 1/r^2$ provided $r\gg\rho^2/2R$,
so the field strength falls off as $1/r^3$ as $r\to\infty$.  In fact, a
more careful study shows that asymptotically the gauge field has the
characteristic of a three-dimensional dipole \gpy.  In the region
$R\ll r\ll \rho^2/2R$ (provided it exists), we find instead that
$A_\mu\sim 1/r$ and the space-time components of the field strength
looks like that of a magnetic monopole.

This correspondence with a magnetic monopole can be made exact when we
set $\rho=\infty$ so that the monopole region above extends to
infinity.  In this limit, the single periodic instanton ansatz,
\singlem, becomes conformally invariant and is in fact gauge
equivalent to the BPS magnetic monopole solution when we identify $A_4$
(the component of the gauge field in the periodic direction) with the
Higgs field in the BPS limit and $1/R$ with the vacuum expectation
value of the Higgs \rossi.  A remarkable aspect of this solution is
that it is actually independent of the periodic coordinate.  This
identification between the periodic instanton and the BPS monopole only
holds for the single monopole case. Anti-monopole solutions are
obtained in this framework by starting with a periodic array of
anti-instantons instead of instantons.

\subsec{Gauge fivebranes and BPS gauge monopoles}

Given the above identification of a single conformal periodic instanton
with a BPS monopole and the close connection between fivebranes and
instantons, it is natural to try to construct monopole solutions in
string theory starting from the fivebrane solutions. Specifically, we
trivially compactify five of the spatial dimensions tangent to the
fivebrane in \fivebr\ on a five-torus and then look for solutions where
in addition the fourth transverse direction is compactified on an $S^1$
of radius $R$.

For the gauge fivebrane the relation is very simple. To construct the
lowest-order single monopole solution of \HL\ we take an array of gauge
fivebranes periodic in the $x^4$ direction.  In particular, we take an
instanton string with spacing $2\pi R = 2\pi/C$ and with size
$\rho=\infty$.  Because the fivebrane energy is independent of $\rho$,
no singularities arise in this limit.  Up to an $x^4$ dependent gauge
transformation, this gauge field configuration is equivalent to the BPS
solution used in \HL, so through \gsol, it yields the identical
solution for the gravity fields given by the dilaton configuration
\eqn\dilgsol{\etp=e^{2\phi_0}+2\apm{1\over r^2}[1-K^2+2H]}
where $H=Cr\coth Cr-1$ and $K=Cr/\sinh Cr$ are the BPS functions.  This
BPS gauge monopole is independent of the internal coordinates and is
thus not only a solution of the compactified ten-dimensional theory but
also of the $N=4$ super Yang-Mills supergravity theory constructed by
dimensional reduction to four dimensions.

Using the ansatz \fivebr\ for the self-dual solution, we deduce that the
non-zero components of the
resulting three-form field strength are given by
%
%
\eqn\hgm{\eqalign{H_{ij4}&=-2\apm\epsilon_{ijk}{x^k\over r^4}H(1-K^2)\cr
&\approx -2\apm C\epsilon_{ijk}{x^k\over r^3}\qquad r\to\infty.\cr}}
Since $H_{\mn 4}$ is the gauge invariant field strength of the $U(1)$
field coming from $B_{\mu 4}$ in the compactification, we see that the
BPS gauge monopole is also an $H_{(4)}$ monopole with magnetic charge
$-8\pi\apm / R$. We will discuss the quantization of the magnetic
$H_{(4)}$ charge in the next section.

When the instanton size is finite, the periodic gauge fivebrane is no
longer independent of the internal direction.  This finite size
instanton solution is equally valid as a solution of the compactified
theory, but it differs from the previous solution in that it cannot be
viewed as a purely four-dimensional solution and it does not have an
interpretation as a non-abelian monopole since the Yang-Mills gauge
field strength falls off like a dipole sufficiently far from the core.
However, it still has the interpretation of a magnetic
$H_{(4)}$ monopole with the same magnetic charge.  We will calculate
the magnetic charge in the next section.

\subsec{Neutral fivebranes and neutral H monopoles}

In a similar fashion, neutral monopole solutions can also be
constructed out of periodic configurations of neutral fivebranes.
Starting with $n$ periodic stacks of neutral fivebranes spaced at a
distance $2\pi R$ apart in the compact dimension, \mnexp\ can be summed
to give
\eqn\multin{\etp(\vx,
x^4)=e^{2 \phi_0}+\sum_{I=1}^n{n_I\apm\over2Rr_I}\sinh{r_I\over R}
\bigg/\left(\cosh{r_I\over R}-\cos{x^4-x^4_I\over R}\right).}
The asymptotic form of the $H$ field can be calculated giving
\eqn\hnm{H_{ij4}
= -\sum_{I=1}^n {n_I\apm \over 4R}\epsilon_{ijk}{(x-x_I)^k\over r_I^3}
+O(e^{-r_I/R})}
for $r_I\gg R$.  Thus, these solutions are multi $H_{(4)}$ monopoles
with total magnetic charge $-\sum_I {n_I\pi \apm / R}$.  Since this
solution is dependent on the internal coordinate $x^4$, it must be
viewed as a compactification of the original 10 dimensional theory.

An $x^4$ independent solution is indicated by considering the formal
limit $R\to 0$. Specifically we want to take this limit while keeping
${n_I\apm / 2R}$ fixed to give
\eqn\nmono{\etp=e^{2\phi_0}+\sum_{I=1}^n{n_I\apm \over 2R}{1\over r_I}.}
There are many reasons why we should be concerned about this limit.
Firstly, because of the quantization of $H$ and hence the discreteness
of $n_I$, the limit isn't properly defined. In addition the limit
involves topology change. However, \nmono\ is indeed a well defined
solution for arbitrary $R$ as can be seen by returning to the derivation
of the neutral
fivebrane \mnexp. Here we want to solve $\sq\,\etp=0$ assuming
$S^2\times S^1$ symmetry and no dependence on the $S^1$ which has
radius $R$. The solution \nmono\ satisfies these conditions and now the
$H$ field given by \hnm\ is valid everywhere for this solution.
The quantization
of the coefficients in \nmono\ comes from the quantization condition on
$H_{(4)}$ that we will discuss in the next section.

The solution \nmono\ was first constructed to lowest order in
\banksetal.  The solution was obtained by a duality transformation of a
solution based on the Sorkin-Gross-Perry monopole. In showing that this
solution can be obtained using the fivebrane ansatz \fivebr, we have
also shown that this solution is a {\it supersymmetric} solution to
string theory.

\subsec{Symmetric fivebranes and symmetric monopoles}

It should now be obvious that one can construct symmetric monopole
solutions from an array of symmetric fivebranes.  Starting with the
periodic instanton configuration \multim, the symmetric monopole
solution is then given by the dilaton field
\eqn\symsol{\etp=e^{2\phi_0} f(\vx,x^4)}
and obviously satisfies the consistency condition $\sq\,\etp=0$.  In
fact, recalling the equivalence of the dilaton fields for the neutral
and symmetric fivebrane solutions, it should be no surprise that this is
the same as for the neutral monopole solution \multin\ after relating the
gauge and gravitational instanton sizes by defining
$\rho_I=e^{-\phi_0}\sqrt{n_I\apm}$.

Similarly, an $x^4$ independent solution is in fact obtained by taking
the limit $R\to0$ and $\rho_I\to0$ with $m_I=\rho_I^2/2R$ fixed.  In
this limit, we find
\eqn\rrkmono{f(x)=e^{-2\phi_0}\etp=1+\sum_{I=1}^n
{m_I\over r_I}}
which reproduces the solution of \khurione. The quantization condition
on $H_{(4)}$ requires that $m_I={e^{-2\phi_0}n_I\apm / 2R}$ and we see
that the dilaton for this symmetric solution is the same as for the
neutral solution \nmono.  We will explain in the next section that
since the Yang-Mills field has a dipole structure at spatial infinity
it cannot be interpreted as a Yang-Mills monopole.

In order to have a true BPS symmetric monopole solution, we instead
take the limit $\rho\to\infty$ for a single periodic symmetric
fivebrane.  Up to an overall rescaling of the metric, this is
equivalent to dropping the 1 in the ansatz \multim.  Thus for a finite
rescaled $\rho$, the BPS symmetric monopole is given by
\eqn\symmono{f(x)=e^{-2\phi_0}\etp={\rho^2\over 2Rr}
\sinh{r\over R}\bigg/\left(\cosh{r\over R}-\cos{x^4\over R}\right).}
Although the $x^4$ dependence of the Yang-Mills field can be gauge
transformed away, the dilaton and gravity fields remain $x^4$
dependent.  In particular, the $x^4$ dependence of the generalized
connection cannot be transformed away since the coordinate is
compact.  Nevertheless, the Bianchi identity is still satisfied since
both $\Tr F\wedge F$ and $\tr R\wedge R$ are $x^4$ independent.

A comparison with the neutral solutions shows that all of these
symmetric solutions are $H_{(4)}$ (multi-)monopoles with unit of
magnetic charge given by $-\pi\apm/R$.

\newsec{Properties of the Solutions}

\subsec{H vs Yang-Mills monopoles}

The compactification from ten to four dimensions introduces several
$U(1)$ gauge fields, six from the $g_{\mu a}$ components of the metric
and six from $B_{\mu a}$ where $a=4,\dots,9$.  As was discussed in
\rohmwitt\ the appropriate gauge invariant field strengths for the
$U(1)$ fields coming from the antisymmetric tensor are given by
$H_{(a)\mn} \equiv H_{a\mn}$.  Of these $U(1)$ gauge fields, only
$H_{(4)}$ is excited in the solutions we have been discussing.

All of the solutions presented in the last section are
$H_{(4)}$ monopole solutions.  In \rohmwitt\ such monopoles were
discussed in a general context and the construction of some solutions
was sketched by postulating the existence of an asymptotic monopole
field strength and then demanding that the field could be smoothly
continued to all of space. Assuming that space had the topology of
$R^3$, it was argued that to avoid Dirac singularities, gauge field
instantons played an important role in the solution via the Bianchi
identity $dH=-{1\over30}\apm \Tr F\wedge F+\ldots$. The gauge monopole
solution, \dilgsol\ and \hgm, is an explicit (and supersymmetric)
realization of this kind of solution. It is interesting to note that
for this solution the $B$ field is not excited and the contributions to
$H_{(4)}$ come entirely from the gauge Chern-Simons term.

In the last section we showed that the BPS gauge monopole solution
\dilgsol\ had $H_{(4)}$ magnetic charge of $-8\pi\apm/R$ by an explicit
calculation of $H_{(4)}$. An alternative way to calculate this and also
to calculate the $H_{(4)}$ charge of the gauge solution with finite
instanton size is to follow an argument presented in \rohmwitt. Using
the fact that the topology of space is $R^3$ and that $H_{(4)}$ is
asymptotically independent of $x^4$, we can relate the $H_{(4)}$ charge to
the instanton number through the use of the Bianchi identity, \bianc,
giving
\eqn\rarg{\eqalign
{g_{(4)}&\equiv \int_{S^2}H_{(4)}={1\over 2\pi R}\int_{S^2\times S^1}H
={1\over 2\pi R}\int_{R^3\times S^1} dH\cr
&=-{\apm\over 2\pi R}\int_{R^3\times S^1}\tr F\wedge F + O(\apm^2)
=-{8\pi\apm q/R}\cr}}
where
\eqn\pont{q={1\over16\pi^2}\int\tr F\wedge F\in Z}
gives the instanton number.  For the BPS gauge monopole, $q=1$, so that
$g_{(4)}=-8\pi\apm/R$.

Since the gauge fields are not excited at all for the neutral monopole
solutions \multin, they do not fit into the scheme discussed in
\rohmwitt.  The reason that the asymptotic monopole field strength can
be continued into the interior of space is that now the topology is not
$R^3$ but $R^3-\{0\}$
as will be discussed in the next subsection. Although an instanton
gauge field is excited for the symmetric monopole solutions,
\symsol\ and \rrkmono, the considerations of the $H$ field is more like
that of the neutral monopoles than that discussed in \rohmwitt.  Note
also that because of the topology of these solutions the $H_{(4)}$
magnetic charge cannot be calculated using \rarg.

We now discuss the Dirac quantization condition for $H_{(4)}$ which
comes from the quantization of the three-form $H$. Specifically,
choosing the manifold $M$ in \hq\ to be an asymptotic $S^2\times S^1$
with $S^2$ at spatial infinity and $S^1$ the compact dimension and
assuming that $H_{(4)}$ is asymptotically independent of the
coordinate on $S^1$ we obtain
\eqn\hfq{\int_{S^2} H_{(4)}=ng\qquad g=-{\pi\apm\over R}}
where $n$ is an integer (positive for the solutions we have been
considering) and $g$ is the unit of magnetic charge.
The states electrically charged with respect to $H_{(4)}$ come from
strings that wind around the $S^1$. To calculate the unit of electric
charge we note that the coupling of the string to the antisymmetric
tensor contains a term
%
\eqn\bfield{S={1\over \pi\apm}\int d^2\sigma\left[\dot X^i\del_\sigma X^4-
\dot X^4\del_\sigma X^i\right]B_{i4}.}
Looking at a configuration that winds around the $S^1$ once, in the
center of mass frame we have a term
\eqn\bfe{S={2R\over\apm}\int dt \dot X^i B_{i4}.}
Thus the winding state couples to the $U(1)$ gauge field like a charged
particle with unit of electric charge given by $e=2R/\apm$.
{}From \hfq\ we see that the charges satisfy the Dirac quantization
condition $g_{(4)}e=2\pi n$.

We now turn to a discussion of the Yang-Mills fields.  Now we are
interested in whether the solutions can be thought of as monopoles of
the $U(1)$ arising from the spontaneous symmetry breaking of $SU(2)$. The
BPS gauge monopole of \HL, constructed out of a periodic gauge
fivebrane, and the BPS symmetric monopole \symmono, constructed out of
a periodic symmetric fivebrane, are obviously of this type.
Asymptotically, the gauge field strength constructed from \symmono\ is
that of a non-abelian magnetic monopole,
$F_{ij}^a\sim-\epsilon_{ijk}x^ax^k/r^4$.  Before gauge transforming,
the ``Higgs field'' behaves as $\varphi^a\equiv A_4^a\sim x^a/r^2$
which has vanishing expectation value.  However, with an $x^4$ dependent
gauge transformation\rossi, $\varphi\to U(\varphi+\partial_4)U^{-1}$
such that $U\partial_4U^{-1}=O(1)$, we recover the BPS solution,
$\varphi^a\sim Cx^a/r$ with the asymptotic $F_{ij}^a$ unchanged.

On the other hand, for the symmetric monopole composed of finite sized
periodic instantons, the non-abelian magnetic field strength,
$F_{ij}^a$, has the characteristics of a dipole and falls off as
$1/r^3$ as $r\to\infty$.  The ``Higgs field'' in this case falls off as
$1/r^2$.  We can again change the asymptotics of $\varphi^a$ by an $x^4$
dependent gauge transformation, but since $F_{ij}^a$ is essentially
unchanged, it still falls off as a dipole. Another way to understand
why these solutions are not non-abelian monopoles is to think about
quantization of the collective coordinates for these solutions. Because
of the falloff of the Higgs and gauge fields at infinity there will be
normalizable zero modes corresponding to global $SU(2)$ gauge rotations
of the solution (in contrast to true non-abelian monopoles for which
these zero modes are not normalizable).  Quantization of the
corresponding collective coordinates will give a spectrum of states in
definite representations of the unbroken gauge group, much as in the
Skyrme model of hadrons.  As a result we conclude that both the
symmetric monopole solutions \symsol\ and \rrkmono\ carry no
non-abelian magnetic charge%
\foot{In this respect, our interpretation differs from that of \khurione.
Without a Higgs vacuum expectation value, the unbroken $U(1)$
electromagnetic field strength tensor of 't~Hooft used in
\khurione\ loses its meaning.}.

\subsec{Spacetime properties of the monopoles}

We now turn to the spacetime properties of the various monopole
solutions.  When investigating these properties, it is important to
keep track of the various possible metrics.  In 10 dimensions, the
above solutions are constructed with the ``sigma model'' metric,
$g_{MN}$.  This is related by a Weyl rescaling,
$g_{MN}=e^{\phi/2}\hat g_{MN}$, to the ``Einstein'' metric
$\hat g_{MN}$ where the action takes the canonical Einstein-Hilbert
form.  When compactified to four dimensions, the contribution of the
internal space volume to the four-dimensional gravitational coupling
can be scaled out to give the four-dimensional canonical metric
$\tilde g_\mn = \Delta^{1/2}\hat g_\mn$ where $\Delta=\det \hat g_{ij}$
(with $\mu,\nu$ being space time and $i,j$ internal space indices).  In
particular, for the fivebrane ansatz, the 10 dimensional sigma model
metric is
\eqn\tenline{ds^2=-dt^2+\etp(dr^2+r^2d\Omega_{(2)}^2+(dx^4)^2)+dx_I^2}
where $x_I$ stands for the remaining 5 internal directions orthogonal
to the fivebrane that can be thought of as being compactified on a five
torus.  The four-dimensional canonical line element has the isotropic
form
\eqn\linee{d\tilde s^2=-e^{-\phi}dt^2+e^{\phi}(dr^2 + r^2 d\Omega_{(2)}^2)}
where the isotropic coordinate $r$ can be related to the Schwarzschild
radial coordinate $\rs$, by $\rs=re^{\phi(r)/2}$.

Examining the four-dimensional metric, \linee, for the BPS gauge
monopole \dilgsol, we find that it has the interesting property that there are
no event horizons or singularities regardless of the Higgs vacuum
expectation value $C=1/R$.  The mass density of the monopole is given
by the 00 component of the stress energy tensor in an orthonormal
basis, $\rho(\rs)=T_{\hat0\hat0}(\rs)$, and the Schwarzschild mass is
$M(\rs)=4\pi\int_0^\rs\rho(\rs)\rs^2d\rs$.  Working in isotropic
coordinates, we calculate for the BPS gauge monopole
\eqn\massd{\rho(r)=T_{\hat0\hat0}(r)=
{\apm e^{-3\phi}\over\kappa_4^2r^4}\left[ 2((1-K^2)^2+2H^2K^2)
+{7\apm e^{-2\phi}\over r^2}H^2(1-K^2)^2\right]}
where the first term is due to the gauge and Higgs field and the second
term is due to the gravity fields.  Here, $\kappa_4$ is the four-dimensional
gravitational coupling.  The mass as a function of the isotropic coordinate
$r$ is most easily calculated from the metric and is given by
\eqn\massM{M(r)={8\pi\apm e^{-3\phi/2}\over\kappa_4^2r}H(1-K^2)\left[
1-{\apm e^{-2\phi}\over2r^2}H(1-K^2)\right].}
In the limit $r\to\infty$, we find the ADM mass of the monopole to be
$M=8\pi\apm C e^{-3\phi_0/2}/\kappa_4^2=-g_{(4)}e^{-3\phi_0/2}/\kappa_4^2$.
The behavior of this monopole is
governed by the dimensionless scale parameter
$\lambda^2=\apm C^2e^{-2\phi_0}$.  For $\lambda\ll1$, the gauge field
dominates, and the monopole has a core size
$r_{\rm core}\approx1.5e^{-\phi_0}\sqrt{\apm}/\lambda$.  The core has a
constant density
$\rho_{\rm core} ={2e^{\phi_0} \over3\apm\kappa_4^2}\lambda^4$ which falls
off as $1/r^4$ outside the core.

For a BPS monopole coupled to gravity, since the mass and inverse size
of the monopole is proportional to $C$, we expect the monopole to
become a black hole when $C$ becomes sufficiently large \josh.  This
fate is avoided in the present situation because of the dilaton field.
For $\lambda \gg 1$, the core mass density still comes from the gauge
field.  However, because of the dilaton coupling, we now find
$\rho_{\rm core}={e^{\phi_0}\over3\sqrt{2}\apm\kappa_4^2}\lambda$, and
$r_{\rm core}\approx2.17e^{-\phi_0}\sqrt{\apm}/\lambda$. The fraction
of the mass concentrated at the core approaches 0 as
$\lambda\to\infty$.  In this limit, the monopole develops an
intermediate region between $r_{\rm core}$ and
$r_{\rm max}\approx10e^{-\phi_0}\sqrt{\apm}\lambda$ in which the mass
density falls off only as $1/r^2$.  The energy in this region arises
from the second term in \massd, and accounts for the majority of the
total mass.

One caveat of this above analysis is that the BPS gauge monopole given by
\dilgsol\ is only valid to lowest order in $\apm$.  The higher order
corrections will presumably be important in the case when $\lambda\gg1$.
Nevertheless, we expect the qualitative behavior to hold for all values
of $\lambda$.  An alternate way to see that the BPS gauge solution is
always well behaved is to note that the underlying gauge fivebrane
solutions are
everywhere regular, regardless of the instanton sizes.

Since the dilaton field has the same form for the neutral and symmetric
solutions they have identical spacetime properties.  For the neutral
and symmetric monopole solutions, \multin\ and \symsol, and at
distances larger than the compactification radius $R$, the
four-dimensional metric approaches \linee\ with
\eqn\asympt{\etp=e^{2\phi_0}\left[1+\sum_{I=1}^n{m_I\over r_I}\right].}
In this limit, the metric is identical to the 4-metric of the
Sorkin-Gross-Perry Kaluza-Klein monopole \refs{\sorkin,\gperry}.  Since
these monopoles are equivalent to a periodic array of neutral or
symmetric fivebranes, their geometry has the structure of the
underlying fivebrane \world.  For distances $r^2+(x^4)^2\ll R^2$, each
monopole core has the geometry of a semi-infinite wormhole in the
original 10 dimensional ``sigma model'' metric \tenline. As mentioned
earlier this ``cylindrical'' topology evades the relation between the
$H$ charge and gauge field instanton charge discussed in \rohmwitt,
essentially by pushing the instanton charge off to the end of the
infinite wormhole.

For the neutral and symmetric solutions, \nmono\ and \rrkmono,  which
describe the limit $R \rightarrow 0$, the asymptotic 4-metric with
\asympt\ is valid everywhere.  In this case, there is a singularity at
the location of the core of each monopole shielded by a horizon located
at $r_i=-m_I$. Since $m_I$ is positive this corresponds to a naked
singularity. It is important to remember here that the anti-monopole
solution is obtained not simply by taking $m_I \rightarrow -m_I$ but
requires going back and starting from an anti-self-dual fivebrane
solution. The relation between the $H$ charge and $m_I$ (or equivalently
$n_I$), \hnm, will
then have an additional minus sign so that the anti-monopole solution
will also have a naked singularity.
Near each monopole core, the metric approaches
\eqn\metrcor{ds^2=-\sqrt{r/me^{2\phi_0}}dt^2+\sqrt{me^{2\phi_0}/r}
(dr^2+r^2d\Omega_{(2)}^2).}
The spatial part of this metric can be converted to Schwarzschild
coordinates according to
\eqn\isom{ds^2=(r/r_0)^s(dr^2+r^2d\Omega_{(d-1)}^2)
={1\over A^2}(d\rs^2+A^2\rs^2d\Omega_{(d-1)}^2)}
where $\rs=r_0(r/r_0)^A$ and $A=1+s/2$.  This metric has a conical singularity
at the origin and in $d$ dimensions has a curvature
$R_{(d)}=(d-1)(d-2)(1-A^2)/\rs^2$ which is non-vanishing for $d>2$.  In this
case,  $s=-1/2$ and the 3-geometry at the origin is that of a conical
space with deficit solid angle $\Omega\equiv4\pi(1-A^2)=7\pi/4$.

What are we to make of these singular solutions? There seem to be two
possibilities. If we truly consider the limit $R \rightarrow 0$ then at
least in the context of string theory we are ignoring the effects of
the light string winding modes which will certainly alter the structure
of the solution. Put another way, as $R \rightarrow 0$ we should be
using a different low-energy effective field theory for the winding
modes rather than the low-energy theory for the momentum modes.  On the
other hand, although we obtained the solutions \nmono\ and \rrkmono\ by
considering the limit $R \rightarrow 0$  we can consider this to be
just a formal trick for obtaining an $x_4$ independent solution and
once we have it, reinstate a non-zero compactification radius $R$.
This is essentially what is done in \khurione.

It thus seems that the symmetric monopole solution \rrkmono\ is
a singular but exact solution to string theory.  It was suggested
in \khurione\ that the divergence in the curvature cancels against the
gauge field singularity, at least in the calculation of the action. It
would be interesting to see whether string theory expanded about this
solution really is non-singular.

The ADM masses of the neutral and symmetric solutions can be read off from
the asymptotic behavior of the metric where the dilaton is given by \asympt.
The result is
$M=(2\pi e^{\phi_0/2}/\kappa_4^2)\sum_I m_I$ which can be rewritten as
$M=(\pi\apm e^{-3\phi_0/2}/\kappa_4^2R)\sum_I n_I
=-g_{(4)}e^{-3\phi_0/2}/\kappa_4^2$.  This relation between the mass and the
$H_{(4)}$ charge is related to the saturation of a Bogomol'nyi bound for
these solitons and holds for {\it all} the above monopole solutions that have
an asymptotically flat metric.

In the case of the BPS symmetric monopole, \symmono, we find
$\etp\sim me^{2\phi_0}/r$ in the $r\to\infty$ limit.  In this case,
neither the original 10 dimensional sigma model metric nor the 4-metric
is asymptotically flat.  The resulting conical metric at infinity is an
indication that the monopole mass is divergent.  This diverging action
is the result of identifying the metric instanton connection with the
gauge connection in the limit of an infinitely large instanton.  Since
the gravity fields do not obey the Yang-Mills equations, they give a
divergent instead of a zero contribution to the (gravitational) energy
density.  In particular, $T_{\hat0\hat0}\sim1/\rs^2$ which is
reminiscent of non-gauge monopoles coupled to gravity\BaVi.  The reason
this is not a problem for the BPS gauge monopole is that in that case
the instanton scale is independent of the scale of the gravity fields.

\newsec{Conclusions}

In this paper we have tried to give a unified description of the known
magnetic monopole solutions to string theory (or its low-energy limit)
by relating them to the three known fivebrane solutions.
We have clarified
the supersymmetry of the solutions which is particularly
significant for the solutions that receive higher order corrections
in $\apm$ since non-renormalization theorems can be used to show that
the corrections will not destabilize the solutions.
We have explained how the
symmetric solutions discussed in \khurione\ are actually limiting cases
of a more general class of symmetric monopole solutions. We have also
emphasized that these symmetric solutions are {\it not} non-abelian
monopoles but are rather monopoles of a $U(1)$ group resulting from
compactification of the antisymmetric tensor field. It would be
interesting to study the analogous dyon solutions, the structure of the
$N=4$ superconformal field theory which underlies the monopole
solutions, and the implications of these solutions for both
$R \rightarrow 1/R$ duality and the more general $SL(2, R)$ duality
recently discussed by Sen in the context of monopole solutions
\sennew.

\bigskip\centerline{\bf Acknowledgements}\nobreak
We would like to thank  Ramzi Khuri, Rafael Sorkin and Andy Strominger
for helpful discussions.  This work was supported in part by the
U.S.~Department of Energy under Grant No.~DE-FG05-85ER-40219 and by NSF
Grant No.~PHY90-00386.  J.P.G.\ is supported by a grant from the
Mathematical Discipline Center of the Department of Mathematics,
University of Chicago. J.H.\ also acknowledges the support of NSF PYI
Grant No.~PHY-9196117.

\listrefs
\end